\documentclass[11pt]{article}
\usepackage[pdftex]{graphicx}
\begin{document}
\title{Software Autotuning for Sustainable Performance Portability}
\author{Azamat Mametjanov and Boyana Norris \\
Mathematics and Computer Science Division\\
Argonne National Laboratory, Argonne, IL 60439\\
\{azamat,norris\}@mcs.anl.gov
}
\date{}                     
\maketitle                  %
\thispagestyle{empty}       
\begin{abstract}
Scientific software applications are increasingly developed by large interdiscplinary teams operating on functional
modules organized around a common software framework, which is capable of integrating new functional capabilities
without modifying the core of the framework. In such environment, software correctness and modularity take precedence at
the expense of code performance, which is an important concern during execution on supercomputing facilities, where
the allocation of core-hours is a valuable resource. To alleviate the performance problems, we propose automated
performance tuning (autotuning) of software to extract the maximum performance on a given hardware platform and to
enable performance portability across heterogeneous hardware platforms. The resulting code remains generic without
committing to a particular software stack and yet is compile-time specializable for maximal sustained performance.
\end{abstract}

\section{Introduction}
Development, testing and debugging of a modern scientific application requires a deep software stack. The
traditional toolchain includes \textit{system} software such as device drivers, operating system and windowing systems,
as well as \textit{development} software such as compilers, debuggers, linkers and virtual machines. Increasingly, there
is a greater reliance on \textit{library} software that provides implementations of common data structures and recurring
algorithms: e.g. BLAS, NetCDF, MPI, Boost among others. Development of a new scientific application involves multiple
design choices at each level of hardware architecture, OS, programming language, compiler and library software stack levels. This
creates a dependency DAG, where dependency edges flow from the lower layers of software stack to the higher level
components. Each vendor of a component in this dependency stream is responsible for ensuring interoperability with
upstream components while providing an API for downstream components. Multiple standard-issuing organizations are
created in this software ecosystem to ensure compatibility and interoperability of components. Some system vendors take
it upon themselves to resolve such issues and create integrated system modules and toolsuites that come along with the
hardware: e.g. IBM compilers and Cray MPI libraries.

While \textit{standardization} and vendor-specific \textit{vertical integration} resolve the issues of syntactic and
semantic component interoperability, performance of a scientific application is subject to substantial variation across
various systems. Large system vendors and leadership-class supercomputing facilities maintain teams of performance
engineers responsible for debugging, deploying, tuning and maintaining scientific applications on large supercomputing
facilities. While a portion of the engineering effort goes toward intrusive source code modifications, some of the
efforts are oriented toward lightweight recognition and fixes of performance affecting patterns. In addition,
performance can depend on the parameters on the underlying platform: e.g. register file size, cache and line size among
many others. Source code of hotspot kernels can be specialized to these parameters to extract the maximal effective
performance. We propose to automate the lower-level performance patterns and specialization using autotuning.

\section{Approach}
The goal of autotuning is to automatically tune the performance of a code for a given platform configuration. Because
the targeted platforms can be heterogeneous, the accompanying benefit of autotuning is performance portability. Tuning
is accomplished by annotating existing code with performance directives in the form of source code pragmas. The
annotation-based approach does not modify the semantics of a given program, which acts as a reference implementation
that can be compiled and executed to obtain reference results. The annotated code is transformed according to
performance directives, compiled and executed to obtain its performance metrics and outputs for comparison with
reference results. Depending on the number of parameter variations within the directives, a number of resulting code
variants are compared and the highest performing variant is selected as the result of autotuning.

\section{Results}
In previous work, we have extended OrioÕs annotation-based approach to generate CUDA GPU code from an existing C code.
Performance exploration parameters include number of threads in a block, number of blocks in a grid of threads,
number of asynchronous streams, size of L1 cache and loop unroll factor among others. The result is an auto-generated
CUDA code that performs better than NVIDIA's cuSPARSE and CUSP library codes
\cite{ChoudaryGHKLMNSS.sisc12,MametjanovLMN.Cluster12}.

\begin{figure}[!htb]
\centering
\includegraphics[width=\linewidth]{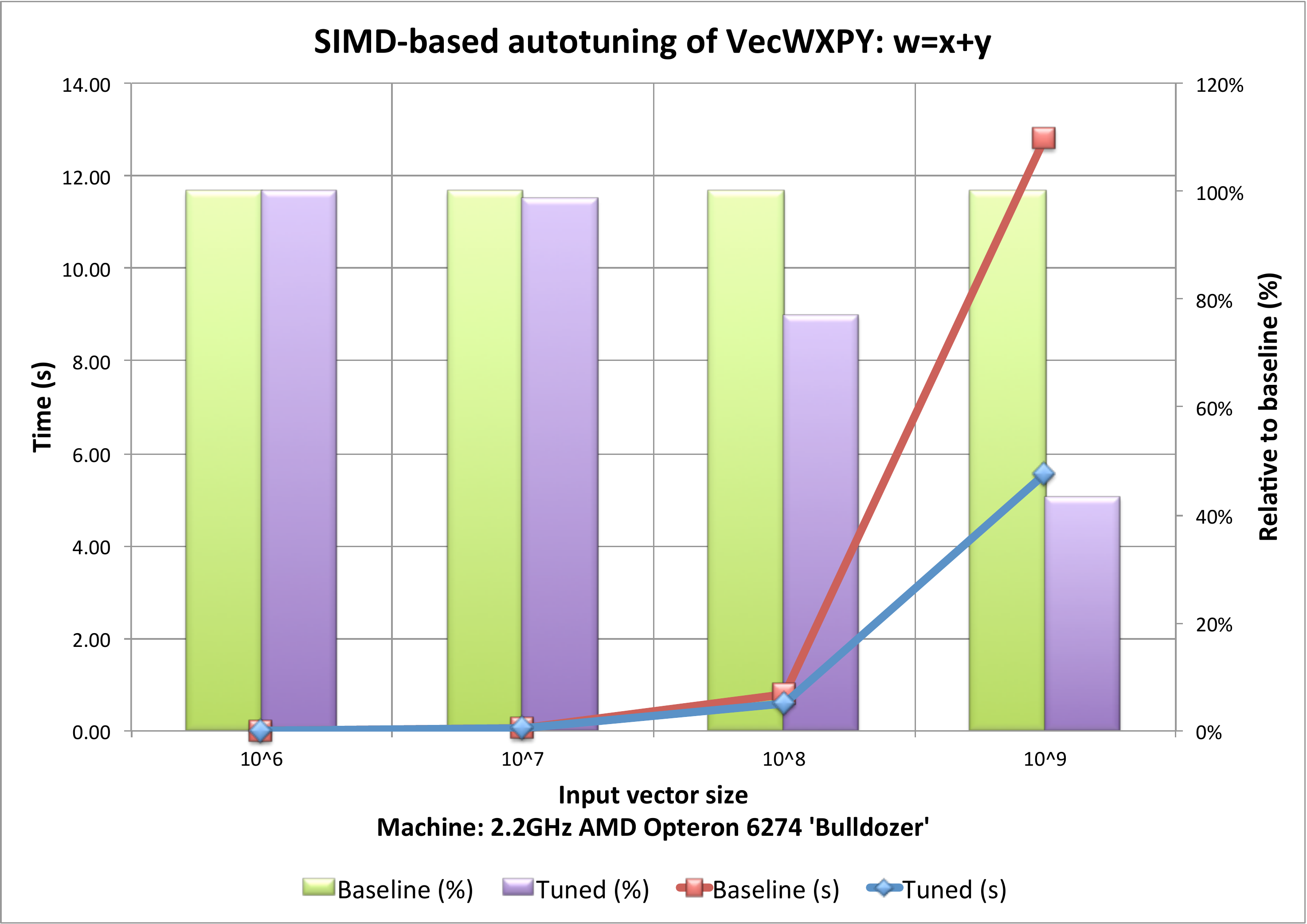}
\caption{Auto-vectorized (baseline) vs. autotuned kernel's performance with Intel's ICC 13.1.3. Each input vector
size is listed with the kernel's absolute execution time (in seconds, lines) along the left vertical axis as well as the
relative speedup of autotuned kernel's time with respect to the baseline kernel's time for that size (in \%\%, bars)
along the right vertical axis. The baseline kernel contains no pragmas and is compiled with '-O3' flag, which turns on
auto-vectorization. Autotuning delivers up to 43\% or 2.3x speedup.}
\label{fig:cd_vecwxpy}
\end{figure}

In the current work, we are using Orio to annotate loops for SIMD vectorization with SSE/AVX. In particular,
instead of relying on a compiler to auto-vectorize the loops, we decorate loops with single-line annotations that
specify a search for SIMD pragmas. Orio transforms the annotated code into a set of compiler-specific pragma-annotated
code variants, guides the compiler in SIMD code generation and selects the highest performing code variant. Figure
\ref{fig:cd_vecwxpy} illustrates initial results of autotuning compared to autovectorizing. 

\section{Conclusion}
In this paper, we have outlined an approach of achieving performance portability across various systems and system
changes. The benefits of this approach are the automation of manual tuning and the specialization of programs to
platforms for better performance. An accompanying benefit is the closure of the gap between theoretical and sustained
performance of scientific applications on modern architectures.

\bibliographystyle{plain}
\bibliography{paper}

\end{document}